# New electronic orderings observed in cobaltates under the influence of misfit periodicities

A. Nicolaou[1,2], V. Brouet[1], M. Zacchigna[3], I. Vobornik[3], A. Tejeda[4], A. Taleb-Ibrahimi[2], P. Le Fèvre[2], F. Bertran[2], C. Chambon[2], S. Kubsky[2], S. Hébert[5], H. Muguerra[5], D. Grebille[5]

[1] *Laboratoire Physique des Solides, Université Paris-Sud, UMR8502, Bât 510, 91405 Orsay, France*
[2] *Synchrotron SOLEIL, L'Orme des Merisiers, Saint-Aubin-BP48, 91192 Gif-sur-Yvette, France*
[3] *CNR - INFM, Lab. Nazionale TASC c/o Area Science Park, s.s. 14 Km. 163.5, I-34012 Basovizza (TS), Italy*
[4] *Institut Jean Lamour, CNRS-Nancy-Université-UPV-Metz, 54506 Vandoeuvre-les-Nancy, France*
[5] *Laboratoire CRISMAT, UMR 6508 CNRS et Ensicaen, 14050 Caen, France*

**Abstract** - We study with ARPES the electronic structure of $CoO_2$ slabs, stacked with rock-salt (RS) layers exhibiting a different (misfit) periodicity. Fermi Surfaces (FS) in phases with different doping and/or periodicities reveal the influence of the RS potential on the electronic structure. We show that these RS potentials are well ordered, even in incommensurate phases, where STM images reveal broad stripes with width as large as 80Å. The anomalous evolution of the FS area at low dopings is consistent with the localization of a fraction of the electrons. We propose that this is a new form of electronic ordering, induced by the potential of the stacked layers (RS or Na in $Na_xCoO_2$) when the FS becomes smaller than the Brillouin Zone of the stacked structure.

In the past decades, many intriguing electronic properties have been discovered in *layered systems* rather than in systems with strictly three dimensional structures. This is often attributed to enhanced quantum effects in low dimensions - two dimensions (2D) for layered systems - that may give rise to original phenomena. In such systems, there is usually a metallic plane (or slab) that is considered as the active unit, stacked with passive structures that mainly act as charge reservoirs. Recently, this view was challenged by an ARPES study of one misfit cobaltate [1], which rather proposes a covalent bonding across the interface between the two types of units. Such questions may also be relevant to address the unexpected properties recently observed at interfaces between two oxides, like superconductivity at the interface between two insulators [2]. Cobaltates offer interesting possibilities to study these questions, because they stack planes with different, sometimes incommensurate, periodicities. It is believed that the influence between the layers could go far beyond a simple transfer of charges [29]. In this paper, we study with ARPES the development of the metallic structure under these two competing potentials in 4 different misfit cobaltates, where both the relative periodicities of stacked planes and the number of transferred electrons between them can be changed. We show that it gives rise to new forms of electronic orderings, characterized by a localization of a fraction of the electrons with structures determined by that of the neighboring plane.

In layered cobaltates, the active unit is a $CoO_2$ slab, formed by a triangular array of Co, embedded in edge-sharing oxygen octahedra. The charge reservoirs are either Na layers in $Na_xCoO_2$ [3] or rock-salt (RS) layers in misfit cobaltates [4]. They may transfer $x$ electrons to the $CoO_2$ slab, yielding an average valence ($4-x$) per Co, hence ($1-x$) holes in the three $t_{2g}$ bands. In Na cobaltates, NMR has revealed that Na may induce partial *charge localization* in the $CoO_2$ planes, with a periodicity related to the Na order. At $x=2/3$, for example, the Co sites located directly underneath a Na trap an electron and become $Co^{3+}$, while the other sites form a metallic kagomé lattice of $Co^{3.5+}$ sites [5]. This is a proof that the two sub-systems cannot be treated as independent. Furthermore, there is an intriguing evolution of the electronic properties, concomitant with the appearance of charge localization at $x>0.6$ [6]. The effective mass increases, strong magnetic correlations develop and the thermoelectric power (TEP) is unusually high for a metal [7,8]. It was proposed that the charge ordering induced by Na could be a source of correlation in this limit [29], but there is not definite understanding of their exact relationship. In this context, it is particularly interesting to study $CoO_2$ slabs in a different 3D environment, as in misfit cobaltates. Our study is organized in three parts. First, we show that ARPES detects replica of the Fermi Surface (FS) with the periodicity of the RS, both in commensurate and incommensurate phases. This implies that the metallic carriers of the $CoO_2$ slabs do sense the RS potential, but this does not require a covalent bonding between the two units, as suggested in [1].

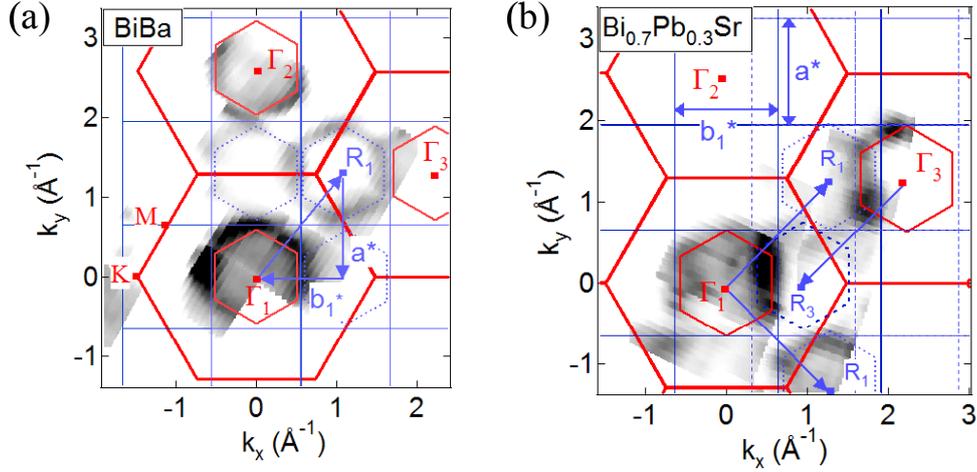

Fig. 1 : FS measured at SIS beamline in BiBa (a) and $Bi_{0.7}Pb_{0.3}Sr$ (b) with 100eV photon energy and a temperature of 20K. Thick red lines correspond to $CoO_2$ BZ and thin blue lines to RS BZ. In $Bi_{0.7}Pb_{0.3}Sr$, where the RS and $CoO_2$ slabs are incommensurate, two sets of RS BZ are shown centered at $\Gamma_1$ (solid blue lines) or $\Gamma_3$ (dotted blue lines).

Second, we detail the form of the RS potential. We show that, surprisingly, it presents strong analogies with that of a Na layer, which suggests the possibility of similar charge localization effects. We further show with STM that, even in incommensurate phases, long range structural orderings are obtained, with super-cells as large as 80Å. Third, we deduce from the FS area the number of metallic carriers in different misfit compounds. We show that it deviates from the expected stoichiometry at high $x$, in a way similar to $Na_xCoO_2$ [9] and consistent with progressive charge localization. Taken together, these three observations support the idea that charge ordering is strongly favored in $CoO_2$ slabs at high $x$ and adopt a structure dictated by the potential of neighboring planes. We finally discuss the influence of these orders on the metallic properties, in relation with the progressive metal-insulator transition observed at high $x$ in misfits [4], but not in Na cobaltates [10].

Our single crystals were prepared by a standard flux method [11] and characterized by structural, magnetic and transport measurements [12]. The general formula of misfit cobaltates is $[(AO)_p(BO)_q]^{RS}.[CoO_2]_m$, abbreviated in the following as AB, where the first bracket describes the ideal composition of the RS block and $m=b_1/b_2$ is the misfit ratio ($b_2=2.82$Å is the side of the triangular lattice, the in-plane RS parameters are $a=\sqrt{3}\ b_2$ and $b_1$, see Fig. 2). The number of electrons in $CoO_2$ planes is defined by the composition of the RS block, but is difficult to estimate a priori, because of frequent vacancies and substitutions in the RS. Bobroff et al. realized that the TEP value at 300K ($S_{300}$) scales with $x$ in $Na_xCoO_2$ and that this scale offers a convenient and reliable way to estimate $x$ also in misfit cobaltates [13]. We adopt this scale to evaluate the doping, giving $x_S=[250+S_{300}]/500$. ARPES measurements were carried out at the SIS beamline of the Swiss Light Source, the APE beamline of ELETTRA and the CASSIOPEE beamline of SOLEIL, with typical energy resolution of 15meV and angular resolution of 0.2°. The STM measurement was performed at the SOLEIL surface laboratory with an Omicron setup.

Fig. 1 presents FS measured in two of the most metallic misfit cobaltates, the nearly commensurate BiBa (m=1.98, $x_S$=0.7) [14] and the strongly incommensurate $(Bi_{0.7}Pb_{0.3})Sr$ (m=1.75, $x_S$=0.71) [12]. We already reported a FS map for BiBa in ref. [4], but, to our knowledge, Fig. 1b is the first report of a FS in an incommensurate misfit cobaltate. The BZ of the triangular $CoO_2$ slab is shown as thick red hexagons and that of the RS planes as thin blue rectangles. In both cases, only the $a_{1g}$ Co band crosses the Fermi Level (FL), giving rise to a hole-like hexagonal FS around $\tilde{\Gamma}$, sketched as thin red line. Similar hexagons are observed in the second $CoO_2$ BZ (centered at $\Gamma_2$ or $\Gamma_3$), meaning that the FS possesses the periodicity of the $CoO_2$ slab. However, Fig. 1 shows that one can further detect *replicas* of these main hexagons, shown as dotted blue hexagons. They are not diffraction replicas, like the replicas generated by the modulation of the Bi-O plane commonly observed in Bi cuprates [15], as their polarization dependence is different from that of the main lines (their periodicity is also completely different from that of the Bi modulation [11]). They indicate a true reconstruction of the electronic structure with the RS periodicity. In BiBa, there is a particularly clear replica centered at $R_1$, which corresponds to the main FS translated by $a^*+b_1^*$. In $(Bi_{0.7}Pb_{0.3})Sr$, the situation becomes more complicated because of the incommensurability that could, in principle, generate an infinity of replicas. Nevertheless, three well defined replicas are observed at the two equivalent $R_1$ points and at $R_3$. They can be obtained by translations of the main FS from $\Gamma_1$ or $\Gamma_3$, respectively, which generate replicas that do not coincide anymore, because of the incommensurability.

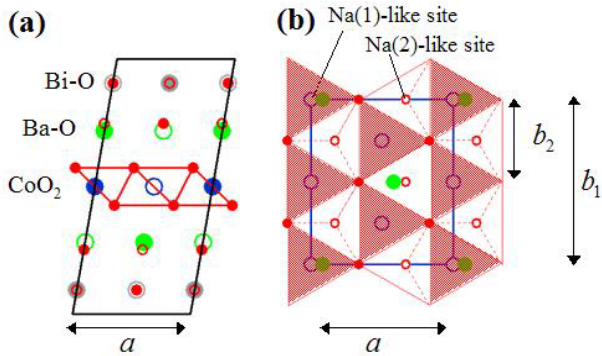

Fig. 2 : (a) 3D stacking of the planes in Bi misfit cobaltates. Closed symbols correspond to b=0 and open symbols to b=1/4 (Co and O) or b=1/2 (Ba and Bi). (b) In-plane projection of the structure. Co atoms are shown as open blue circles, Ba above the $CoO_2$ plane as green filled circles and oxygens above (below) the $CoO_2$ plane as small solid (open) red circles. Oxygen triangles above a Co ion are shown as dashed red.

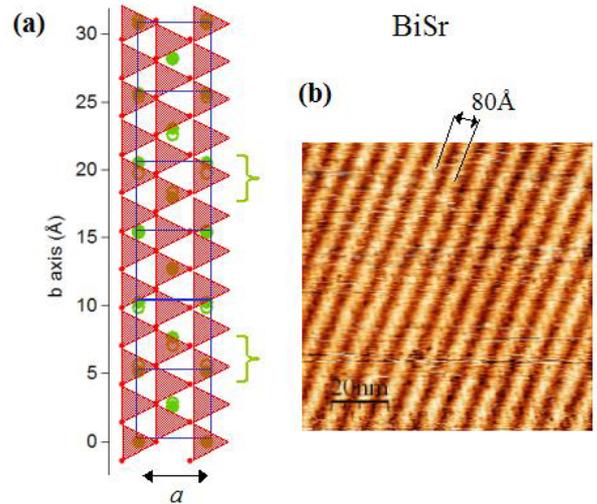

Fig. 3 : (a) In-plane structure for BiSr over 6 neighboring RS cells (blue rectangles). Green points represent Sr atoms above the $CoO_2$ slab and open green circles the nearest stable position above an oxygen triangle (small red points). (b) STM image of BiSr at 300K. The stripes are aligned along a. This image was acquired with a voltage of 3.5V and a feedback current of 1.5nA. It was also reproduced in different experimental conditions, for three different samples.

Replicas are expected to be observed in ARPES, every time there is a new periodicity in a system [16], whether it originates from neighboring layers or is intrinsic to one layer, as for example in a Charge Density Wave (CDW). These two types of replica were for example clearly observed in the CDW compound $CeTe_3$ [17]. In principle, their intensity is a measure of the strength of the coupling at the origin of the new periodicity [16] and it decreases fast for high order replicas. The observation of replicas in misfit cobaltates is therefore not so surprising, but proves that the coupling between the $CoO_2$ and RS planes is large enough to affect the electronic structure. Unfortunately, it is not possible to be more quantitative, as the ARPES intensities are strongly modulated by matrix-element effects. One can observe in Fig. 1 that the intensities of the different replicas are quite different (this is also true for the main FS in different BZ) and we observed that they also depend strongly on photon energy.

Similar replicas were recently reported by Ou *et al.* in BiBa [1], but they were attributed to an intrinsic metallic character of the RS planes. As stated in introduction, this would force to reconsider considerably the usual understanding of doping in layered systems. We believe that the clear hexagonal shape of the replica in Fig. 1a, which was not clearly resolved in the work of Ou *et al.*, is an unambiguous proof that it belongs to the $CoO_2$ electronic structure, and not to the RS. The impression that the main peaks and replicas originate from two different sub-systems was enhanced in their work by the fact that they observed either the main peaks or their replicas, but this is just due to a peculiar matrix element effects [18] and we show here that they usually coexist. We also show here that replica are a general feature of misfit cobaltates that is not associated to the strain of the RS, contrary to assumptions of Ou *et al.* (this strain is indeed much reduced in $(Bi_{0.7}Pb_{0.3})Sr$, where $a \approx b_1$). Therefore, we do not see any reason to assume metallicity of the RS planes to describe these observations, which just attest that the carriers, likely confined in the $CoO_2$ slabs, do sense the RS periodicity.

We may now turn to the second part of our investigation, which aims at understanding the *nature* of the coupling detected between the RS and $CoO_2$ slabs. The main reason for the carriers to feel the RS periodicity is that the RS creates a potential that adds to the one of the $CoO_2$ slab. One could first expect this potential to be quite smooth, especially compared to that of a Na layer, as there is a regular alternation of $A^{2+}$ and $O^{2-}$ ions. However, a closer look at the actual structure in Fig. 2a shows that there is a strong displacement of 0.4Å of the A ions towards the $CoO_2$ planes. This is a common feature of all misfit compounds [1,11,14]. This will create a net positive charge, as seen from the $CoO_2$ plane, which resembles the case of a $Na^+$. Moreover, Fig. 2b reveals that, in BiBa, the sites for $Ba^{2+}$ are nearly the same as that of $Na^+$. Na locates preferentially at the center of an oxygen triangle, either just above a Co [this site is called Na(1)] or in the middle of a Co triangle [called Na(2)] [3]. In BiBa, the Ba at the corners of the cell occupies a position similar to Na(1) and those at the center a position similar to Na(2). The only difference is that, whereas Na is free to move exactly to the center of the oxygen triangle, the Ba positions are primarily constrained to a rectangular lattice. As the centers of the triangles are not regularly spaced, Ba moves to an average position, by shifting the whole plane by 0.4Å in the a direction (this is the origin of the tilt of the whole structure represented in Fig. 2a). For the Ba-O plane below the $CoO_2$ slab, the positions of the Ba are just translated by $b_1/2$. Note that this configuration is exactly that of $Na_{0.5}CoO_2$, although the doping is different in BiBa. In $Na_{0.5}CoO_2$, it has often been argued that this particular Na order could induce a charge segregation between rows of $Co^{3+}$ and $Co^{4+}$, explaining the metal-insulator transition specific to this compound [7] (it was later showed that the charge segregation, if present, must have a much smaller amplitude [19]). Our description of the BiBa structure reveals that a similar influence of the RS potential on electronic orderings in the $CoO_2$ slabs could also be discussed in misfit structures. We will return to this point later.

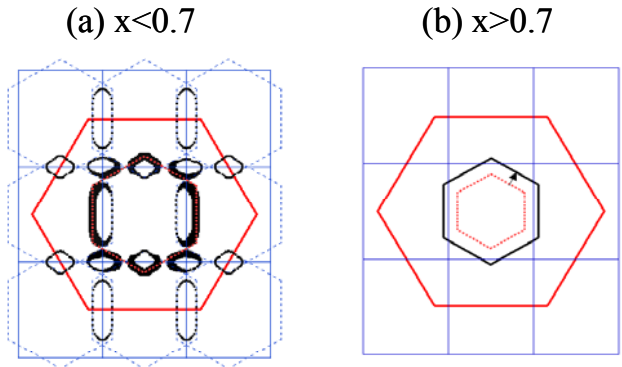

Fig. 4 : Sketch of the different FS behaviors expected as $x$ increases. The red solid hexagon is the $CoO_2$ BZ, the blue rectangle the RS BZ and the dotted red hexagon, the FS before reconstruction. The FS after reconstruction is shown with black lines. (a) The FS is larger than the RS BZ. A gap opens when the FS crosses its replicas (dotted blue hexagons), which lowers the electronic energy and creates small FS pockets. The thickness of the lines is proportional to the ARPES spectral weight, which remains concentrated along the original hexagon [16]. (b) The FS is smaller than the RS BZ. No gap would open at the Fermi level in these conditions, unless the FS is enlarged by localization of a fraction of the charges (black hexagon). This FS could also undergo reconstruction, which is not sketched.

For misfit phases, the evolution of the RS potential with the incommensurability is particularly interesting. When $b_1$ shrinks, the density of $A^{2+}$ ions per Co increases, quite similarly as when $x$ increases in the Na layer. This also moves the $A^{2+}$ charges away from the stable positions above the oxygen triangles, until a coincidence is recovered for some common multiple of $b_1$ and $b_2$. This is sketched in Fig. 3a for the case of BiSr (m=1.82 and $x_S$=0.75), where six RS unit-cells define an approximately commensurate unit-cell ($6b_1$=10.9$b_2$=31Å). Of course, the position of the $A^{2+}$ charges may distort to optimise their location above oxygen triangles. Indeed, structural modulations are commonly observed in misfits, both in A-O and Bi-O layers [11,20]. However, Fig. 3a shows that a simple shift along b to the nearest stable position results in regions (indicated by large brackets), where Sr atoms would be too close from each others. Therefore, a more complicated modulation, probably also involving shifts along the z axis, should be found. These modulations will inevitably create a strong modulation of the $A^{2+}$ potential sensed by Co along b, which raises in turn the question of its consequences on the distribution of charge carriers in the plane.

Interestingly, the STM image of BiSr in Fig. 3b reveals an extremely clear pattern of "stripes", ordered over very large areas (at least 1μm*1μm). The stripes are aligned along a and exhibit a periodicity of 80Å along b, much larger than the approximately commensurate super-cell considered in Fig. 3a. The periodicity and direction of the stripes do not match that of the Bi modulation (q*=0.293a*+0.915c* in BiSr, i.e. 17Å along a [11]). This pattern indicates that the system finds a way to reorganize over surprisingly large unit cells and that the RS potential is certainly not random or disordered at the local scale. If interpreted topologically, the difference in height between dark and bright regions is about 1Å, which could well produce large differences of potentials attracting preferentially, and maybe pinning, charges in one of these regions. We do not know at present whether this ordering is present throughout the whole bulk or enhanced at the surface and whether there is an electronic ordering associated to this pattern. Nevertheless, this shows that the carriers of the $CoO_2$ slabs in misfit cobaltates will be submitted to a RS potential with an amplitude modulated along b with a periodicity ranging from the unit cell (BiBa) up to very large unit-cells (BiSr), depending on m.

Two different effects of the RS potential could be considered. The first one is the reconstruction of the electronic structure with the RS periodicity, evidenced by the presence of replicas. If the FS and its replicas intersect, this also leads to the FS recontruction sketched in Fig. 4a. Gaps open at the crossings between the original FS and its replicas, leading to small FS pockets. Such a mechanism was proposed to explain the de Haas-Van Alphen oscillations observed in Na cobaltates, which correspond to very small FS areas [27]. As the ARPES spectral weight remains concentrated along the original hexagon [16], such a reconstruction will be more difficult to detect with ARPES, especially if the gaps are small or the order is not perfect. Such a picture then allows to reconcile the large hexagonal FS observed with ARPES and the small FS pockets observed with de Haas-Van Alphen measurements.

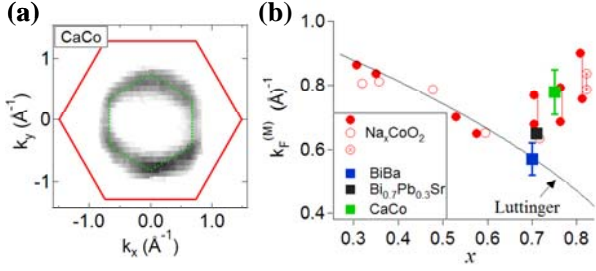

Fig. 5 : (a) FS in CaCo measured at the CASSIOPEE beamline, at 50K and 100eV photon energy. (b) Circles : $k_F$ along ΓM as a function of $x$ in $Na_xCoO_2$ taken from ref. 9 (solid), ref. 20 (open) and our work (crossed). For $x>0.7$, the FS splits and we report the two $k_F$ values. Squares : $k_F^{(M)}$ in different misfits as a function of $x_S$, the equivalent $x$ calculated from the TEP scale (see text).

Another possible consequence of the RS potential could be to *localize* holes or electrons at minima or maxima of this potential, as seems to occur in $Na_xCoO_2$ [5,28,29]. This will change the effective number of carriers in the metallic band, therefore the size of the unreconstructed FS, as sketched in Fig. 4b. The effective number of metallic carriers in the different phases can be deduced from the area of the unreconstructed FS using the Luttinger theorem [21]. Assuming a hexagonal FS, this area just depends on one $k_F$ value, for example $k_F^{(M)}$ along ΓM, and the number of metallic holes is $n_h=2(k_F^{(M)}/ΓK)^2$ (the factor 2 accounts for spin degeneracy). This corresponds to an effective doping $x_m=1-n_h$. In BiBa, $k_F^{(M)}=0.57(5)$Å$^{-1}$ was determined from measurements in 4 different samples, including Fig. 1a and the more accurate measurement of ref. [4]. It yields $x_m=0.7$, in perfect agreement with the TEP scale $x_s$. In $(Bi_{0.7}Pb_{0.3})Sr$, $k_F^{(M)}$ is less clearly defined because of the overlap of the replicas. The hexagon of Fig. 1b is drawn with $k_F^{(M)}=0.65$Å$^{-1}$. In BiSr, the QP is unfortunately too small to extract reliably a $k_F$ value, although FL crossings also clearly exist there. An additional measurement is presented in Fig. 5a for CaCo, a misfit with only 3 RS layers (Ca-O/Co-O/Ca-O), m=1.62 and $x_S=0.75$ [22]. We confirmed it in 3 different samples, yielding $k_F^{(M)}=0.78(7)$Å. The fact that the FS is larger in CaCo than in BiBa is then clearly above experimental incertitude. However, this means that $x_m=0.45$ in CaCo in strong disagreement with the TEP scale and all other properties of CaCo, which resemble that of cobaltates with $x>0.7$ : the susceptibility becomes Curie-Weiss, it acquires negative magnetoresistance and orders magnetically below T=30K [22,23]. It is also less metallic than BiBa [19], like other misfit cobaltates with $x>0.7$ [12].

This apparent contradiction may in fact reveal very important physics of the $CoO_2$ slabs. We plot in Fig. 5b the values of $k_F^{(M)}$ obtained in misfits, together with those obtained by previous studies of $Na_xCoO_2$ [9,24], as well as our own observation in $Na_{0.82}CoO_2$. The solid line describes the evolution of $x_m$ with $k_F^{(M)}$ expected from the Luttinger theorem. We observe that, although the Luttinger theorem is well obeyed for $x<0.7$ [24], *there is a strong deviation for $x>0.7$ in both families*. The deviation then appears robust in the $CoO_2$ plane and cannot be easily attributed to a surface problem [25], the magnetic state or the dimensionality [9], which are all quite different in misfit and Na cobaltates.

An interesting point is that this deviation appears above $x=0.6-0.7$, precisely in the region where Na cobaltates cease to be conventional metals, but exhibit magnetic correlations [7] and static $Co^{3+}$ [5,6]. As a first approximation, these $Co^{3+}$ can be treated as disordered, just taking electrons out of the metallic band, resulting in more holes in the metallic band than expected from stoichiometry and larger FS. An effective band filling $x_m=0.45$ in CaCo would for example be consistent with $x_S=0.75$, if 55% of the Co sites are localized $Co^{3+}$. A tendency to electron localization then explains quite well the deviation and even seems to us an important confirmation of this phenomena in $Na_xCoO_2$ at $x>0.6$ (this was also suggested in ref. [26]), but also in misfit cobaltates, which it is rather unexpected. Electron localization then appears as an intrinsic tendency of the $CoO_2$ plane in this doping range. Note that the deviation rules out *hole* localization, i.e. localized $Co^{4+}$ magnetic moments, which was proposed to explain the Curie-Weiss susceptibilities [27], but would, on the contrary, decrease the FS area [30]. The information on the RS potential gathered in this paper led us to conclude that the RS potential plays the role of Na potential in misfit cobaltates. We note that if the $Co^{3+}$ adopt an ordered structure, as seems to be the case for Na cobaltates, at least in the bulk [5], a new "metallic" BZ should be defined and the application of the Luttinger theorem becomes more complicated. However, the qualitative tendency to increase the FS area remains.

This study then implies an extreme sensitivity of the electronic structure of the $CoO_2$ slab to the potential of neighboring layers (NL) at high $x$. As calculated for $Na_xCoO_2$ [28], this is reasonable as these potentials easily exceed the very small QP energy scale probed in ARPES measurements [4,9,24]. The fact that it occurs only in a limited doping range and quite reproducibly in two families that are very different from the structural point of view further implies some intrinsic connection between the electronic structure and this charge localization. Indeed, we can expect two different regimes as a function of doping for the coupling to the NL potential, as sketched in Fig. 4. At low $x$, the FS is large and will, in most cases, intersect the NL BZ (we consider here the elementary NL BZ, such as the blue rectangle ($a^*,b_1^*$) of Fig. 1 for misfits, which is the basic additional periodicity modulating the electronic structure). We then expect a reconstruction that effectively lowers the electronic energy (Fig. 4a).. At higher $x$, the FS shrinks and should not intersect the NL BZ anymore. Fig. 1 shows that this regime is just achieved for BiBa, suggesting a transition for $x \approx 0.7$. In this regime, no electronic energy would be gained anymore by gap openings, unless charge localization enlarges back the FS to produce intersections. Indeed, $x \approx 0.7$ corresponds well to the doping at which the deviation from Luttinger theorem starts. The system may gain energy by reconstructing the enlarged FS but also by choosing a structure for the localized charges that optimises the Coulombic interactions with the NL. We propose that this is the reason for stabilizing such orders, whose originality resides in the fact that the periodicity cannot be adapted to the electronic structure (e.g. $k_F$ in a CDW) but is fixed by an external potential. Different orders are probably realized in the two cobaltate families. Localized $Co^{3+}$ are known to form in $Na_xCoO_2$ under Na(1) sites, leaving a 2D metallic lattice, different from the triangular one. In misfits, we rather anticipate a type of 1D ordering in broad "stripes" with modulated charges, as suggested by the STM pattern in BiSr. Note that the modulation does not necessarily create $Co^{3+}$, which were not directly detected by NMR in this compound [13]. These different orders in turn affect the electronic properties : $Na_xCoO_2$ remain good metals up to high $x$ [10], while misfit cobaltates become almost insulating [4]. This could mean that the 2D charge order in $Na_xCoO_2$ let the itinerant carriers flow, while the 1D order of misfit cobaltates act as barriers. In any case, the fine structure of the FS becomes much more complicated than expected and this is probably an important fact to take into account to describe the intriguing evolution of the electronic properties in this limit.

We would like to thank H. Alloul and J. Bobroff for many useful discussions, L. Patthey for his help at the SIS beamline, D. Roditchev and T. Cren for additional STM measurements and T. Mentes for complementary LEEM experiments on BiSr. AN acknowledges support from the EC contract No. MEST CT 2004 514307 EMERGENT CONDMAT-PHYS Orsay.